\documentclass[superscriptaddress,aps,prl,twocolumn]{revtex4-2}

\usepackage{bm}
\usepackage{dcolumn}
\usepackage{graphicx}
\usepackage[colorlinks=true, linkcolor=blue, citecolor=blue, urlcolor=blue]{hyperref}
\usepackage{mathrsfs}
\usepackage{multirow}
\usepackage{rotating}
\usepackage{url}
\usepackage{color}
\usepackage{amsmath}
\usepackage{amssymb}
\begin{document}


\title{Electrical Regulation of Transverse Spin Currents \\in Unconventional Magnetic Ferroeletrics}

\author{Yudi Yang}
\thanks{These authors contributed equally to this work.}
\affiliation{Zhejiang University, Hangzhou, Zhejiang 310058, China}
\affiliation{Department of Physics, 
Westlake University, Hangzhou 310030, Zhejiang, China}
\affiliation{Institute of Natural Sciences, Westlake Institute for Advanced Study, Hangzhou 310024, Zhejiang, China}

\author{Zhuang Qian}
\thanks{These authors contributed equally to this work.}
\affiliation{Institute for Theoretical Sciences, Westlake Institute for Advanced Study, Westlake University, Hangzhou 310024, Zhejiang, China}

\author{Ruichun Xiao}
\affiliation{Institute of Physical Science and Information Technology, Anhui University, Hefei 230601, Anhui, China}
\affiliation{Anhui Provincial Key Laboratory of Magnetic Functional Materials and Devices, School of Materials Science and Engineering, Anhui University, Hefei 230601, Anhui, China}

\author{Yuanyuan Xu}
\affiliation{Center for Quantum Matter, School of Physics, Zhejiang University, Hangzhou 310058, Zhejiang, China}

\author{Hua Wang}
\affiliation{Center for Quantum Matter, School of Physics, Zhejiang University, Hangzhou 310058, Zhejiang, China}

\author{Shi Liu}
\email{liushi@westlake.edu.cn}
\affiliation{Department of Physics, 
Westlake University, Hangzhou 310030, Zhejiang, China}
\affiliation{Institute of Natural Sciences, Westlake Institute for Advanced Study, Hangzhou 310024, Zhejiang, China}

\author{Congjun Wu}
\email{wucongjun@westlake.edu.cn}
\affiliation{Institute of Natural Sciences, Westlake Institute for Advanced Study, Hangzhou 310024, Zhejiang, China}
\affiliation{Institute for Theoretical Sciences, Westlake Institute for Advanced Study, Westlake University, Hangzhou 310024, Zhejiang, China}
\affiliation{New Cornerstone Science Laboratory, Department of Physics, 
Westlake University, Hangzhou 310024, Zhejiang, China}
\date{\today}

\begin{abstract}
We identify hexagonal YMnO$_3$ as a material realization of the elusive $\beta$-phase of unconventional magnetism, a noncollinear, noncoplanar antiferromagnetic state defined by intrinsic spin-momentum locking and a topological spin texture. First-principle calculations reveal that this unique electronic structure enables a perpendicular electric field to generate a transverse pure spin current, a response that occurs without requiring relativistic spin-orbit coupling. Symmetry analysis demonstrates that this spin current is intimately related to the material's ferroelectric polarization that breaks the inversion symmetry and is rigorously forbidden at domain walls where electrical polarization vanishes.  This provides a blueprint for a non-volatile transistor where a gate voltage switches the spin current conductivity by controlling domain wall density, enabling all-electrical control for energy-efficient antiferromagnetic spintronics.
\end{abstract}
\maketitle
\newpage

A primary goal of spintronics is the generation and control of spin-polarized currents, for which ferromagnetic (FM) materials have traditionally played a central role. 
However, the stray magnetic fields they generate
pose challenges for high-density device integration. 
Although stray fields are absent in conventional collinear antiferromagnets (AFM), these materials typically lack the spin-split band structures essential for generating and manipulating spin currents. 
Recently, altermagnetism (AM) has emerged as a new magnetic phase offering a potential solution to these limitations~\cite{Smejkal22p040501, Smejkal22p031042, Song25p473}. 
Defined as a distinct collinear AFM state, it features spin-split band structures despite having zero net magnetization. 
The underlying symmetry of these materials is described by spin groups, which treat spatial and spin symmetry operations independently~\cite{Xiao24p031037, Chen24p031038, Jiang24p031039}. 
Spin group symmetry operations impose stricter constraints on the allowed elements
of response tensors than magnetic group~\cite{Chen24p031038}. 
Recently, the anomalous Hall effect (AHE) has been reinterpreted as a spin-group symmetry-breaking phenomenon, where the main driving factor is the spin-orbit interaction strength~\cite{Liu25p031006}.

\begin{figure*}
\includegraphics[width=0.8\textwidth]{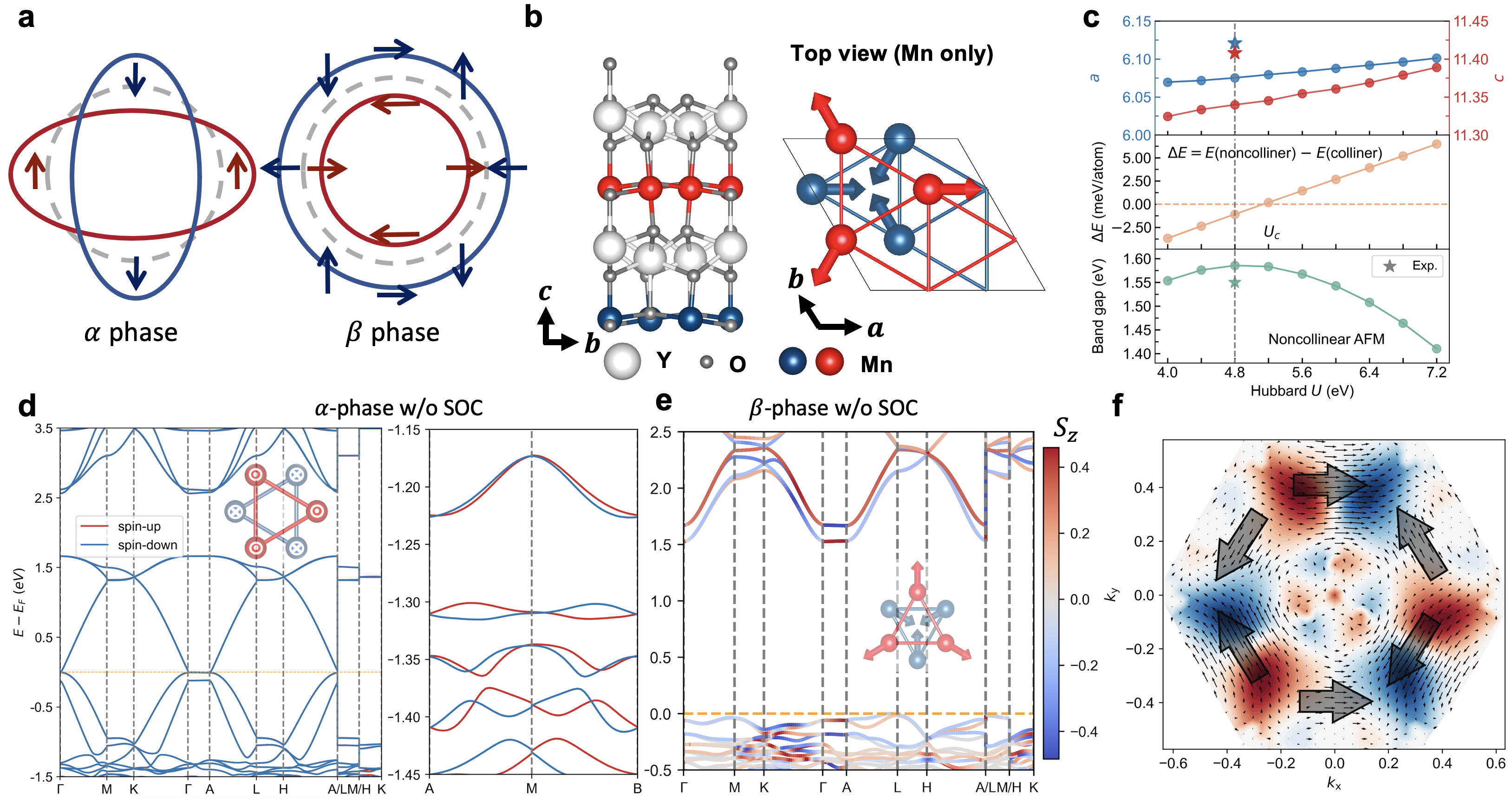}
\caption{\textbf{Electronic structure and intrinsic spin-momentum locking in YMnO$_3$ with $\beta$-phase unconventional magnetism.} 
(a) Schematic top view of the Fermi surfaces in $k$-space for the $\alpha$- and $\beta$-phases, 
respectively, both exhibiting the $l=2$ symmetry in two dimensions.  
(b) Atomic structure of the hexagonal YMnO$_3$ shown from the front and top views.  
(c) Benchmarking of computational parameters. A Hubbard $U$ of 4.8 eV reproduces the experimental lattice constant and correctly predicts the noncollinear, noncoplanar AFM ground state over the collinear state, $\Delta E = E{\mathrm{(noncollinear)}} - E{(\mathrm{collinear})} < 0$, while also capturing the experimental band gap. The bottom panel shows the band gap of $\beta$-phase remains robust across a wide range of $U$ values, even beyond $U_c$.
(d) Band structure of the collinear AFM $\alpha$-phase, showing spontaneous spin splitting without relativistic spin-orbit coupling (w/o SOC).  
(e) Band structure of the noncollinear, noncoplanar AFM $\beta$-phase, colored by the expectation value of $S_z$. Both (d) and (e) are calculated without SOC.  
(f) Spin texture in the $k_x$–$k_y$ plane at $k_z = 0$ for the $\beta$-phase with an energy window of $-0.01$ to $-0.09$ eV.
}
\label{StruEngBand}
\end{figure*}

The recent focus of altermagnetism \cite{Smejkal22p040501, Smejkal22p031042} fits naturally into the broader framework of 
``unconventional magnetism" (UM), which was proposed to spontaneously generate spin-momentum-locking via electron-electron interactions ~\cite{Wu04p036403, Wu07p115103,Lee2009}. 
Applicable to correlated systems in the non-relativistic limit, UM provides the novel mechanism of the emergence of effective spin-orbit coupling (SOC) via quantum phase transitions, rather than by relativity. 
UM is classified into two primary types, the anisotropic $\alpha$-phase and
isotropic $\beta$-phase, in analogy to the superfluid $^3$He A and B-phases, respectively.
Figure~\ref{StruEngBand}(a) illustrates the Fermi surfaces for both phases with an orbital angular momentum number $l=2$. 
The $\alpha$-phase, characterized by anisotropic Fermi surface splitting, shares the same symmetry as altermagnetism for even-$l$ states, indicating they can be smoothly connected through adiabatic evolution.  
An earlier work by four of the authors has proposed that RuO$_2$ can be viewed as an $\alpha$-phase of UM ~\cite{Qian25p174425}.

In contrast, the $\beta$-phase of UM is characterized by an isotropic Fermi surface splitting and a momentum-space spin texture defined by a nontrivial topological winding number $w = l$, where $l$ denotes the partial wave quantum number. In real space, this momentum-space spin texture manifests as a distinct noncollinear AFM ground state. 
Such noncollinear magnetic phases have recently garnered significant interest due to their ability to host rich emergent spin-transport phenomena~\cite{Kimata19p627, Busch21p184423, Rimmler24p109, HAN25p100012, Uchimura25p096701}. 
However, it has not been widely recognized that many of these systems can be fundamentally understood as material realizations of the UM $\beta$-phase, which offers a unifying theoretical framework rooted in many-body interactions. A key feature of these systems is a unique mechanism for spin-current generation that permits both time-reversal-even ($\mathscr{T}$-even) and time-reversal-odd ($\mathscr{T}$-odd) contributions, even in the absence of SOC~\cite{Zhang18p073028, Uchimura25p096701, Zhang17p075128, Jakub17p187204, Rimmler24p109, HAN25p100012, Gurung21p124411, Rimmler24p109}. The $\mathscr{T}$-even component originates from the band topology, whereas the $\mathscr{T}$-odd component arises from scattering-induced relaxation processes~\cite{Freimuth14p174423}.  
In the earlier studies of UM, the $l=2$ $\beta$-phase was already analyzed to generate spin current 
in response to an applied electric field 
via a rank-3 tensor \cite{Wu07p115103}.
Specifically, our analysis shows that an electric field applied along the $\hat x$-direction could result in a pure spin current flowing along $\hat y$ that is polarized along $\hat y$, establishing a directional locking between the flow orientation and spin polarization (see Fig.~S1). 

Despite its intriguing symmetry properties, a definitive material realization of the $\beta$-phase has not yet been  conclusively demonstrated. 
Identifying and characterizing such a material would potentially open new avenues to 
quantum materials with unconventional magnetic and transport properties.
In this work, we establish hexagonal multiferroic YMnO$_3$ as a model system for the elusive 
$\beta$-phase. 
Through a combined analysis incorporating spin group symmetries and first-principle calculations, we demonstrate that noncollinear, noncoplanar AFM YMnO$_3$ exhibits a distinctive form of spin-current directional locking, a hallmark feature of the $\beta$-phase. 
In this state, the spin polarization is rigidly coupled to the spin current direction and simultaneously exhibits both $\mathscr{T}$-odd and $\mathscr{T}$-even components. 
Moreover, the spin current conductivity can be manipulated via the intrinsic ferroelectricity of YMnO$_3$, enabling a direct coupling between electric and spin degrees of freedom. Leveraging this interplay, we propose a non-volatile spintronic transistor that achieves all-electric control of spin currents by manipulating ferroelectric domain walls. 

The hexagonal manganite YMnO$_3$ is a prototypical multiferroic where improper ferroelectricity and frustrated magnetism are intimately coupled to the crystal structure. 
Below the ferroelectric Curie temperature of 1270~K, a structural trimerization, driven by buckling of the MnO$_5$ bipyramids and cooperative displacements of Y$^{3+}$ ions, breaks inversion symmetry and induces a spontaneous electric polarization along the $c$-axis~\cite{ukaszewicz74p81}.  As shown in Fig.~\ref{StruEngBand}(b), the magnetic order originates from Mn$^{3+}$ ions arranged in two-dimensional triangular lattices within each layer. When viewed along the $c$-axis, the Mn sublattice forms a bilayer structure within the unit cell, with alternating red and blue layers exhibiting a staggered, noncollinear AFM configuration~\cite{Fiebig02p818, Fiebig00p5620, Kim24p243}. 
While this 120$^\circ$ noncollinear, noncoplanar spin configuration is experimentally established, accurately reproducing it with density functional theory (DFT) has been a challenge. 
Our analysis shows that the multi-orbital Hubbard parameters of intra-orbital repulsion $U=8.0$ eV and Hund's coupling $J=0.88$ eV
widely adopted in prior studies~\cite{Fennie05p100103, Prikockyt11p214301} incorrectly favor a collinear AFM state over the experimentally observed noncollinear configuration [Fig.\ref{StruEngBand}(c)]. To establish a physically consistent model, we performed a systematic calibration of the Hubbard $U$ parameter.
We find that an optimal value of $U = 4.8$ eV successfully stabilizes the noncollinear, noncoplanar AFM ordering as the ground state. This value also yields lattice parameters ($a = 6.08$\AA, $c = 11.34$\AA) and a band gap (1.59 eV) in good agreement with experimental measurements ($a = 6.121$\AA, $c = 11.408$\AA~\cite{Lee08p7180}; band gap $\approx 1.55$ eV~\cite{Degenhardt01p139}).

\begin{table}
\renewcommand{\arraystretch}{1.}
\caption{Response tensor for polar and nonpolar phases of YMnO$_3$.
Spin current conductivity tensors $\sigma^{S_\alpha}_{ij}$ for polar YMnO$_3$ (space group $P6_3cm$) are derived based on spin group symmetries (indicated in brackets within the table), using \texttt{TensorSymmetry}~\cite{xiao26p109872}. The tensor follow the relation the relation $J_i^\alpha = \sigma^{S_\alpha}_{ij} E_j$, where $J_i^\alpha$ is the spin current flowing in direction $i$ with spin polarization along $\alpha$, and $E_j$ is the applied electric field along direction $j$. Here, $\alpha$, $i$, and $j \in {x, y, z}$ denote Cartesian directions.  In the polar phase, the response tensors are identical for both $\mathscr{T}$-odd and $\mathscr{T}$-even components. For comparison, the corresponding symmetry-constrained tensor forms for the nonpolar phase (space group $P6_3/mmc$) are also provided.
}
\begin{tabular}{c | c c c } 
\hline 
\hline
 & \multicolumn{1}{c}{State I} &
 & \multicolumn{1}{c}{Nonpolar} 
 \\
 & \multicolumn{1}{c}{$(P^{3^1_{001}}3^{2_{100}}c^11)$} &
 & \multicolumn{1}{c}{$(P^{-6^1_{001}}6_3/^{1}m^{2_{100}}c^{m_\frac{\pi}{6}}m)$} 
 \\
 \hline
& \multicolumn{3}{c}{$\mathscr{T}$-even}
\\
\hline
$\sigma^{S_x}$ 
& $\begin{pmatrix}\ \ \ \sigma^{S_x}_{xx}\ \ \ & 0 & \ \ 0 \ \ \ \ \\ 
0 & -\sigma^{S_x}_{xx} & \sigma^{S_x}_{yz} \\ 
0 & \ \ \sigma^{S_x}_{zy} & 0 \end{pmatrix}$ 
&
& $\begin{pmatrix} \ \ \ 0\ \ \  &\ \ \  0 \ \ \ & \ \ \ 0 \ \ \ \\ 
 \ \ \ 0 \ \ \  & \ \ \  0  \ \ \ & \ \ \  0 \ \ \  \\ 
 \ \ \ 0 \ \ \  & \ \ \  0 \ \ \  & \ \ \  0 \ \ \  \end{pmatrix}$ \\
$\sigma^{S_y}$ 
& $\begin{pmatrix}\ \ \ \ 0\ \ \ \ \ & -\sigma^{S_x}_{xx} & -\sigma^{S_x}_{yz} \\ 
-\sigma^{S_x}_{xx} & 0 & 0 \\ 
-\sigma^{S_x}_{zy} & 0 & 0 \end{pmatrix}$ 
&
& $\begin{pmatrix} \ \ \ 0\ \ \  &\ \ \  0 \ \ \ & \ \ \ 0 \ \ \ \\ 
 \ \ \ 0 \ \ \  & \ \ \  0  \ \ \ & \ \ \  0 \ \ \  \\ 
 \ \ \ 0 \ \ \  & \ \ \  0 \ \ \  & \ \ \  0 \ \ \  \end{pmatrix}$ \\
$\sigma^{S_z}$ 
& $\begin{pmatrix}\ 0\ \ \ \ \ & \sigma^{S_z}_{xy} \ \ \ \ & \ \ 0 \ \ \ \\ 
-\sigma^{S_z}_{xy} \ \ \ \ & 0 \ \ \ \ &\ \ 0\  \ \ \\ 
\ 0\ \ \ \ \ & 0 \ \ \ \  &\ \ 0 \ \ \ 
\end{pmatrix}$ 
&
& $\begin{pmatrix}  \ \ \ 0 \ \ \ \ & \sigma^{S_z}_{xy} &  \ \ \ 0 \ \ \ \ \\ 
-\sigma^{S_z}_{xy} & 0 & 0 \\ 
0 & 0 & 0 \end{pmatrix}$ \\
\hline 
& \multicolumn{3}{c}{$\mathscr{T}$-odd}
 \\
\hline
$\sigma^{S_x}$ 
& $\begin{pmatrix}\ \ \ \sigma^{S_x}_{xx}\ \ \ & 0 & \ \ 0 \ \ \ \ \\ 
0 & -\sigma^{S_x}_{xx} & \sigma^{S_x}_{yz} \\ 
0 & \ \ \sigma^{S_x}_{zy} & 0 \end{pmatrix}$ 
&
& $\begin{pmatrix} \sigma^{S_x}_{xx}  &\ \ \  0 \ \ \ & \ \ \ 0 \ \ \ \\ 
 \ \ \ 0 \ \ \  & -\sigma^{S_x}_{xx} & \ \ \  0 \ \ \  \\ 
 \ \ \ 0 \ \ \  & \ \ \  0 \ \ \  & \ \ \  0 \ \ \  \end{pmatrix}$ \\
$\sigma^{S_y}$ 
& $\begin{pmatrix}\ \ \ \ 0\ \ \ \ \ & -\sigma^{S_x}_{xx} & -\sigma^{S_x}_{yz} \\ 
-\sigma^{S_x}_{xx} & 0 & 0 \\ 
-\sigma^{S_x}_{zy} & 0 & 0 \end{pmatrix}$ 
&
& $\begin{pmatrix}  \ \ \ 0 \ \ \ \ & \sigma^{S_x}_{xx} &  \ \ \ 0 \ \ \ \ \\ 
-\sigma^{S_x}_{xx} & 0 & 0 \\ 
0 & 0 & 0 \end{pmatrix}$ \\
$\sigma^{S_z}$ 
& $\begin{pmatrix}\ 0\ \ \ \ \ & \sigma^{S_z}_{xy} \ \ \ \ & \ \ 0 \ \ \ \\ 
-\sigma^{S_z}_{xy} \ \ \ \ & 0 \ \ \ \ &\ \ 0\  \ \ \\ 
\ 0\ \ \ \ \ & 0 \ \ \ \  &\ \ 0 \ \ \ 
\end{pmatrix}$ 
&
& $\begin{pmatrix} \ \ \ 0\ \ \  &\ \ \  0 \ \ \ & \ \ \ 0 \ \ \ \\ 
 \ \ \ 0 \ \ \  & \ \ \  0  \ \ \ & \ \ \  0 \ \ \  \\ 
 \ \ \ 0 \ \ \  & \ \ \  0 \ \ \  & \ \ \  0 \ \ \  \end{pmatrix}$ \\
\hline 
\hline
\end{tabular}
\label{TensorTable}
\end{table}

Our benchmark analysis of the Hubbard $U$ parameter can be interpreted as a theoretical model for correlation-driven phase transition in YMnO$_3$ between two distinct UM phases: an insulating $\beta$-phase and a metallic, altermagnetic $\alpha$-phase. We identify a critical value, $U_c \approx 5.2$ eV, which marks the crossover point in the energetic stability of these two competing states [Fig.~\ref{StruEngBand}(c)].
Below this critical value ($U < U_c$), the system energetically favors the experimentally consistent noncollinear AFM configuration, realizing the insulating $\beta$-phase [Fig.\ref{StruEngBand}(e)]. Above the critical value ($U > U_c$), the ground state switches to a collinear AFM order, corresponding to the metallic $\alpha$-phase. This higher-$U$ phase is characterized by $l=4$ ($g$-wave) spin multipoles and the spin-group symmetry $[C_2 \parallel C_{6z}t]$, which gives rise to a prominent altermagnetic band splitting [Fig.~\ref{StruEngBand}(d)].

We find that the band gap of the $\beta$-phase exhibits exceptional robustness, remaining nearly constant approximately at $1.55$ eV across a wide range of $U$ values (4.0–7.2 eV), even into the regime where it is no longer the lowest-energy state [Fig.~\ref{StruEngBand}(c)]. 
This insensitivity demonstrates that the insulating behavior is an intrinsic consequence of the noncollinear spin geometry itself, with the magnitude of the correlation strength playing a secondary role to the spin texture in opening the gap.
The $\beta$-phase characterized by $l=2$ spin-quadrupole moments is expected to possess a topologically nontrivial spin texture in momentum space, where spins at opposite momenta are parallel [Fig.~\ref{StruEngBand}(a)]. Our DFT calculations confirm this topology, revealing a winding number of $w=2$ for the spin texture within an energy window of $-0.01$ to $-0.09$ eV below the valence band maximum [Figs.~\ref{StruEngBand}(f) and S2(a)]. The spin texture of the $\beta$-phase could be accessed experimentally through chemical doping. 
These results suggest that the transition at $U_c \approx 5.2$ eV can be viewed as a correlation-driven first-order UM phase transition from a $g$-wave ($l=4$) $\alpha$-phase to a spin-quadrupole ($l=2$) $\beta$-phase.

\begin{figure}
\includegraphics[width=0.45\textwidth]{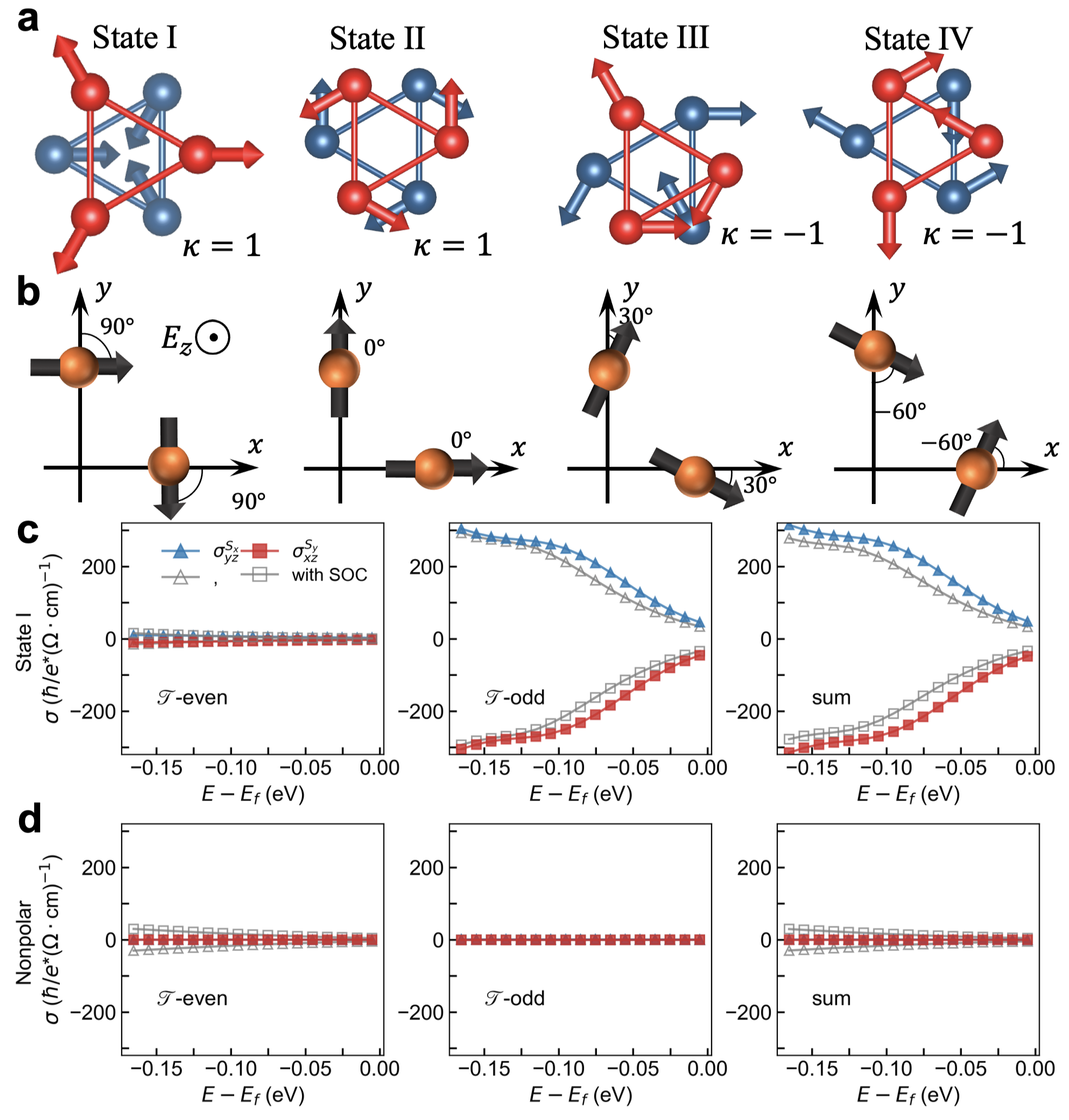}
    \caption{\textbf{Chirality-dependent spin current response in YMnO$_3$ with $\beta$-phase UM}. (a) Four distinct magnetic chiral states, characterized by a net chirality $\kappa = \pm 1$. (b) Schematic illustration of the decomposition of the pure spin current into $x$- and $y$-direction components under an applied electric field along the $z$-axis ($E_z$).(c) Calculated spin-current conductivity ($\sigma_{yz}^{S_x}$ and $\sigma_{xz}^{S_y}$) for the State I compare with (d) the nonpolar phase, where spin current generation is suppressed.
}
\label{SHESSE}
\end{figure}

The noncollinear magnetic order depicted in Fig.~\ref{StruEngBand}(b) is referred to as State I in this study. 
This state is one of four possible configurations we investigate, which are derived from the irreducible representation of the crystalline $P6_3cm$ space group
\cite{Munoz00p9498, Lorenz13p497073}. 
As illustrated in Fig.~\ref{SHESSE}(a), these four states are defined to explore the effects of magnetic chirality: States I and II possess a chirality of $\kappa=+1$, while their chiral partners, States III and IV, have the opposite chirality of $\kappa=-1$. In the absence of SOC, these four states are energetically degenerate. Crucially, our spin texture analysis confirms that, despite their opposite real-space chiralities, all four magnetic states belong to the $\beta$-phase, as evidenced by a common winding number of $w = 2$ in their momentum-space spin textures [Fig.~S3(b)].

While the four degenerate noncollinear magnetic states 
(I--IV) of the YMnO$_3$ are difficult to resolve with conventional probes, we propose that their distinct pure spin current responses provide a direct method for their identification. 
Although these configurations share the same momentum-space topological properties, their real-space spin group symmetries are different.  
This difference in symmetry 
results in the unique form of the spin current response tensor for each state. Guided by this principle, we compare the response tensors for states I-IV, as well as the nonpolar state that exists above the ferroelectric Curie temperature. 
Table~\ref{TensorTable} details the tensor for State I and contrasts it with that of the nonpolar state. 
The spin current response tensors with in-plane spin polarizations ($\sigma^{S_x}$ and $\sigma^{S_y}$) are symmetric, while the response tensor for spin polarization along the $z$-direction ($\sigma^{S_z}$) is antisymmetric. Furthermore, the $\mathscr{T}$-even and $\mathscr{T}$-odd tensor elements share the same symmetry properties. They originate intrinsically from the Berry curvature and from dissipative relaxation, respectively. 
We specifically focus on transverse spin currents generated by an electric field applied along the crystallographic $c$-axis ($E_z$), which corresponds to the third column of each spin current response tensor.
For state I, the  $E_z$ field generates an in-plane spin current with a purely transverse spin polarization: a current flowing in the $y$-direction is $x$-polarized ($\sigma^{S_x}_{yz}$), and a current in the $x$-direction is $y$-polarized ($\sigma^{S_y}_{xz}$). 

The relation between current direction and spin polarization is distinct for each magnetic configuration, yet can be unified within a single framework. 
The in-plane spin current ($j_i^\mu$) generated by $E_z$ is characterized by both the direction of current flow ($i =x, y$) and the spin polarization axis ($\mu = x, y$). This response is captured by the following matrix equation:
\begin{eqnarray}
j^\mu_i (\kappa, \theta) = N E_z
\left(
\begin{array}{cc}
\cos \theta     & \kappa \sin \theta \\
-\sin \theta     & \kappa \cos \theta
\end{array}
\right),
\label{eq:spincurrent}
\end{eqnarray}
where $\kappa$ is the chirality number, $N$ is a normalization constant, and $\theta$ is a state-dependent phase angle. Particularly, the spin current flowing the direction of the azimuthal angle of $\varphi$ is polarized along the direction of $\kappa (\varphi +\theta)$. 
For States I, II, III, and IV, the angle $\theta$ takes the values $-\pi/2$, $0$, $\pi/6$, and $-\pi/3$, respectively. These distinct values of $(\kappa, \theta)$ lead to unique spin current patterns for each state, as depicted in Fig.~\ref{SHESSE}(b).
Therefore, by applying an electric field along the $c$-axis and measuring the direction-dependent spin accumulation on the sample boundaries, one can unambiguously identify the underlying magnetic ground state configuration.

To provide a verification of the above symmetry analysis, we perform first-principle calculations of the spin current conductivity. 
In a magnetic material like YMnO$_3$ where $\mathscr{T}$ is broken, the spin conductivity tensor is composed of two distinct parts: a $\mathscr{T}$-even (intrinsic, $\sigma_{ij}^{\rm{e}}$) component and a $\mathscr{T}$-odd (dissipative, $\sigma_{ij}^{\rm{o}}$) component. Using a tight-binding Hamiltonian derived from our WANNIER90~\cite{Mostofi08p685} results, we computed both components with the Kubo formalism~\cite{Freimuth14p174423, Qiao18p214402}:
\begin{align}
&\sigma_{ij}^{\rm{o}} 
    = -\frac{e \hbar \Gamma^2}{\pi V} 
    \sum_{k,n,m} \frac{ \text{Re}(\langle nk|\hat{j}^{S_\alpha}_i|mk\rangle \langle mk|\hat{v}_j|nk\rangle)}{w(\epsilon_{nk},\Gamma )w(\epsilon_{mk},\Gamma )}, \\
&\sigma_{ij}^{\rm{e}} = -\frac{2e\hbar }{V}  \sum^{\substack{\rm{n\ occ} \\ \rm{m\ unocc}}}_{k,n\neq m} \frac{\text{Im}(\langle nk|\hat{j}^{S_\alpha}_i|mk\rangle \langle mk|\hat{v}_j|nk\rangle)}{(\epsilon_{nk} - \epsilon_{mk})^2}.
\end{align}
The term $w(x,\Gamma)=(x-E_F)^2 + \Gamma^2$ with $\Gamma$  represents a Lorentzian broadening function, where $\Gamma$ is the energy broadening parameter,  
$\epsilon_{nk}$ is the band energy for band $n$ at wave vector $k$,  
and $E_F$ is the Fermi energy. 
The spin-current operator is defined as $\hat{j}^{_\alpha}_i = \frac{1}{2}\{S_{\alpha}, \hat{v}_i\}$ and $\hat{v_i}=\frac{\partial H}{\hbar \partial k_i}$ is the velocity operator. 
Our calculations are carried out by systematically varying the Fermi energy within the valence band. A shift of $E_F$ by 0.07 eV corresponds to a hole doping level of 0.5 holes per unit cell (see Fig.~S4). A broadening parameter $\Gamma = 0.05$ eV is adopted. Throughout this analysis, we adopt the rigid band approximation, assuming that the underlying electronic structure remains unchanged by doping. 

As shown in Fig.~\ref{SHESSE}(c), the spin current response in State I of YMnO$_3$ is governed by its dissipative, $\mathscr{T}$-odd component, which exceeds the $\mathscr{T}$-even component by an order of magnitude. 
This dominance demonstrates that the often-overlooked $\mathscr{T}$-odd contribution is crucial for a correct qualitative description of the material's transport properties. Furthermore, our calculated conductivity tensors for both components fully satisfy the symmetry constraints predicted by spin group analysis ($\sigma^{S_x}_{yz} =-\sigma^{S_y}_{xz} $). We confirmed that SOC, which primarily affects the weak $\mathscr{T}$-even component, is negligible in YMnO$_3$.

We find that the in-plane spin current response components ($\sigma_{xz}$ and $\sigma_{yz}$), generated by $E_z$, are highly sensitive to the ferroelectric polarization ($P$). Our symmetry-based tensor analysis (Table~\ref{TensorTable}) shows that these components are strictly forbidden in the nonpolar state ($P = 0$). A detailed derivation is provided in Supplementary Information (SI). This prediction is directly confirmed by our DFT calculations without SOC, which yield vanishing spin conductivities, as shown in Fig.~\ref{SHESSE}(d). When SOC is included, a finite but weak $\mathscr{T}$-even contribution emerges, yet the total spin current remains at least an order of magnitude smaller than that in the ferroelectric phase.
This pronounced contrast in spin conductivity between the polar and nonpolar states provides a clear mechanism for electrical regulation. In particular, 180$^\circ$ ferroelectric domain walls separate regions of opposite polarization and necessarily pass through an intermediate nonpolar ($P = 0$) configuration. These domain walls are inactive for spin current generation and can therefore serve as ``OFF" channels, enabling a reconfigurable spintronic device architecture based on ferroelectric domain patterning.

\begin{figure}
\includegraphics[width=0.45\textwidth]{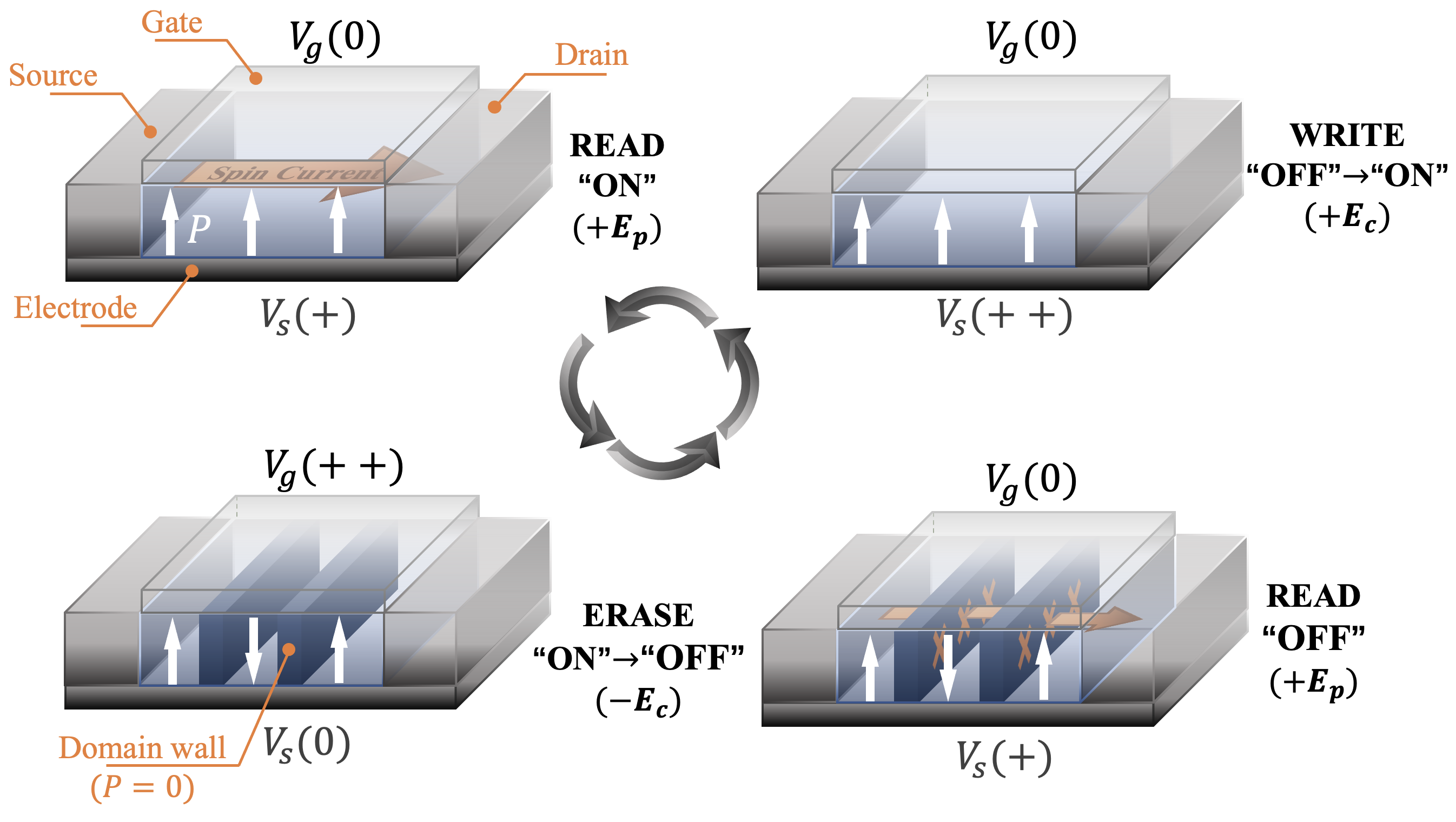}
\caption{\textbf{Nonvolatile spintronic transistor based on $\beta$-phase UM.}
The channel material consists of YMnO$_3$, hosting the noncollinear, noncoplanar AFM state I depicted in Fig.~\ref{SHESSE}(a).
The operation cycle begins in the \textbf{READ ``ON"} state (top left), where a weak, non-switching probing field ($+E_p$), applied via a small voltage $V_s(+)$ to the uniformly polarized channel, generates a maximal spin current along the $x$-direction ($\sigma_{xz}^{S_y}\gg0$).
The \textbf{ERASE ON$\rightarrow$OFF} step (bottom left) is triggered by applying a strong negative gate voltage $V_g(++)$, producing a large electric field ($-E_c$) that partially reverses ferroelectric domains and introduces a high density of domain walls, where the local polarization vanishes ($P = 0$).  
In the subsequent \textbf{READ OFF} state (bottom right), the same weak probing field yields a suppressed spin current, as domain walls are inactive for spin current generation ($\sigma_{xz}^{S_y} \approx 0$).  
Finally, the \textbf{WRITE OFF$\rightarrow$ON} operation (top right) applies a strong positive field to remove the domain walls, restoring the single-domain high-conductance state and completing the nonvolatile spintronic memory cycle.
}
    \label{Transistor}
\end{figure}

Based on this principle, we propose a novel transistor design utilizing a YMnO$_3$ thin film as the active channel between source and drain electrodes (Fig.~\ref{Transistor}). The fundamental operational mechanism is the electrical control of 180$^\circ$~domain wall density within the channel, which directly modulates the in-plane spin current.
In a uniformly polarized state ($|P| > 0$), the channel supports a large spin current in response to $E_z$, as discussed earlier. We note that the spin current generation is insensitive to the direction of polarization, as both $+P$ and $-P$ states yield high spin conductivity. Thus, any single-domain polarized state functions as a high-conductance ``ON" state.
Conversely, applying an electric field of opposite polarity can nucleate a high density of 180$^\circ$ domain walls within the channel. These walls locally reduce polarization to $P = 0$, thereby suppressing spin current generation. This defines the ``OFF" state of the device.

The device's function as a non-volatile memory element is realized through a full WRITE-READ cycle (Fig.~\ref{Transistor}).  
The READ ``ON" operation applies a weak, non-switching probe field ($+E_p$) to the uniformly polarized channel to generate a maximal spin current.
The EARSE operation is then executed by applying a strong coercive field of opposite polarity ($-E_c$), which populates the channel with a high density of domain walls. Importantly, the voltage duration is controlled to prevent full polarization reversal, resulting in a multi-domain configuration with significant $P=0$ regions, where spin current generation is suppressed.  
To confirm the transition, a subsequent READ ``OFF" step uses the same weak probe field ($+E_p$), which now yields a strongly reduced spin current due to the presence of inactive domain walls. Finally, the cycle is completed with a WRITE ``ON" operation, where a large positive field ($+E_c$) sweeps the domain walls from the channel, restores the single-domain, uniformly polarized state, and returns the device to its initial high-conductance condition.

In conclusion, our study identifies noncollinear magnetic YMnO$_3$ as a representative of the $\beta$-phase within the framework of unconventional magnetism, offering a unique platform for spintronics. We demonstrate that its noncollinear, noncoplanar antiferromagnetic state exhibits intrinsic spin-momentum locking, enabling a transverse spin current driven purely by an electric field, without requiring spin-orbit coupling. Symmetry analysis and first-principle calculations reveal that this response is critically dependent on the presence of ferroelectric polarization ($P$) and is strictly suppressed when $P = 0$. Leveraging this principle, we propose a transistor concept in which ferroelectric domain walls act as electrically controlled ``OFF" switches for the spin current. This enables all-electrical, nonvolatile control, where a gate voltage toggles spin transport by creating or removing domain walls. Our results provide a blueprint for energy-efficient memory and logic devices based on electrically tunable spin transport in systems with unconventional magnetism.

\section{acknowledgments}
We gratefully acknowledge Yutong Yu from Southern University of Science and Technology and Junjie Kang from the University of Toronto for helpful conversations. 
C. W. is supported by the National Natural Science Foundation of China (NSFC) under the Grant No. 12234016 and also supported by the NSFC under the Grant Nos. 12174317. 
S.~L. is supported by the Zhejiang Provincial Natural Science Foundation of China (LR25A040004). 
R. C. X. is supported by NSFC under Grant Nos. 12474100 and 12204009.
H. W. is supported by the NSFC under Grant Nos. 12522411, 12474240, and 12304049.
This work has been supported by the New Cornerstone Science Foundation.
The computational resource is provided by the Westlake HPC Center.

\bibliography{note.bib}
\end{document}